\documentclass[%
 reprint,
superscriptaddress,
 amsmath,amssymb,
 aps,
]{revtex4-2}

\usepackage{graphicx}% Include figure files
\usepackage{dcolumn}% Align table columns on decimal point
\usepackage{bm}% bold math
\usepackage[separate-uncertainty = true,multi-part-units=single]{siunitx}
\usepackage[version=4]{mhchem}
\usepackage{tikz}
\usepackage{lipsum}
\usepackage{braket}
\begin{document}

\preprint{APS/123-QED}

\title{Modeling Quantum Volume Using Randomized Benchmarking of Room-Temperature NV Center Quantum Registers}

\author{Tom Jäger}
\affiliation{3rd Institute of Physics, University of Stuttgart, Stuttgart, 70569, Germany}

\author{MinSik Kwon}
\affiliation{3rd Institute of Physics, University of Stuttgart, Stuttgart, 70569, Germany}
\affiliation{Max Planck Institute for Solid State Research, Stuttgart, 70569, Germany}

\author{Max Keller}
\affiliation{3rd Institute of Physics, University of Stuttgart, Stuttgart, 70569, Germany}

\author{Rouven Maier}
\affiliation{3rd Institute of Physics, University of Stuttgart, Stuttgart, 70569, Germany}
\affiliation{Max Planck Institute for Solid State Research, Stuttgart, 70569, Germany}

\author{Nicholas Bronn}
\affiliation{IBM Quantum IBM T.J. Watson Research Center, Yorktown Heights, NY, 10958, USA}

\author{Regina Finsterhoelzl}
\affiliation{Department of Physics, University of Konstanz, Konstanz, 78464, Germany}

\author{Guido Burkard}
\affiliation{Department of Physics, University of Konstanz, Konstanz, 78464, Germany}

\author{Leon Büttner}
\affiliation{Fraunhofer Institute for applied Solid State Research IFA, Freiburg, 79108, Germany}

\author{Rebekka Eberle}
\affiliation{Fraunhofer Institute for applied Solid State Research IFA, Freiburg, 79108, Germany}

\author{Daniel Hähnel}
\affiliation{Fraunhofer Institute for applied Solid State Research IFA, Freiburg, 79108, Germany}

\author{Vadim Vorobyov}
\affiliation{3rd Institute of Physics, University of Stuttgart, Stuttgart, 70569, Germany}

\author{Jörg Wrachtrup}
\affiliation{3rd Institute of Physics, University of Stuttgart, Stuttgart, 70569, Germany}
\affiliation{Max Planck Institute for Solid State Research, Stuttgart, 70569, Germany}

\date{\today}

\begin{abstract}
Accurately estimating the performance of quantum hardware is crucial for comparing different platforms and predicting the performance and feasibility of quantum algorithms and applications.
In this paper, we tackle the problem of benchmarking a quantum register based on the NV center in diamond operating at room temperature. 
We define the connectivity map as well as single qubit performance. 
Thanks to an all-to-all connectivity the 2 and 3 qubit gates performance is promising and competitive among other platforms.
We experimentally calibrate the error model for the register and use it to estimate the quantum volume, a metric used for quantifying the quantum computational capabilities of the register, of 8. 
Our results pave the way towards the unification of different architectures of quantum hardware and evaluation of the joint metrics. 
\end{abstract}
\maketitle
\section{Introduction}
Over the last decades, quantum computing successfully evolved from single-qubit systems to systems with up to hundreds of qubits, which are represented by many quantum technological platforms \cite{vazquez2024scalingquantumcomputingdynamic}.
Among these are cold atoms \cite{bluvstein2022quantum}, superconducting qubits \cite{google2023suppressing,devoret2013superconducting}, trapped ions \cite{leibfried2003quantum,cirac1995quantum}, quantum dots \cite{burkard,buch2013spin, donnelly2024noise}, defects in solids \cite{van2024mapping}, and photons \cite{o2009photonic}.
Each platform has its individual advantages, which naturally arise from the physical nature of the system. 
For example, the superconducting qubits being macroscopic objects can be manufactured and scaled using highly developed chip technology, while trapped atoms benefit from their identical and fundamentally scalable atomic structure. 
Solid-state spin systems typically show long coherence times, high degrees of inter-connectivity, and easier integration into existing solid-state systems. 
Keeping in mind these highly diverse platforms for quantum computing, it is essential to identify the strengths and weaknesses of each platform, as well as to find a unique comparative metric.
For operations performed on a small number of qubits, randomized benchmarking \cite{rb_of_gates, scalable_rb} constitutes such a unifying metric free from initialization and readout errors. 
It allows to identify the error per gate for all single- and two-qubit gates in a register or a chip and to establish an error model. 
Here we perform a randomized benchmark of single- and two-qubit gates in a state of the art four-qubit quantum register based on the Nitrogen Vacancy (NV) center in diamond operating at room temperature to extract the error rates per gate.
Based on our results, we build an error model of the register, which allows the simulation of its performance for various quantum algorithms and reveals avenues for further optimization of the realization of the algorithms. 
Also, we determine the quantum volume of the system, which in our case is limited by the number of addressable qubits as well as the longitudinal relaxation of the central spin at room temperature. 
Our approach can be generalized to registers operating at lower temperatures, where the longitudinal relaxation will not limit coherence and a much higher quantum volume can be achieved.
\section{The NV center at room-temperature}
\begin{figure*}[ht]
    \centering
    \includegraphics[width=\textwidth]{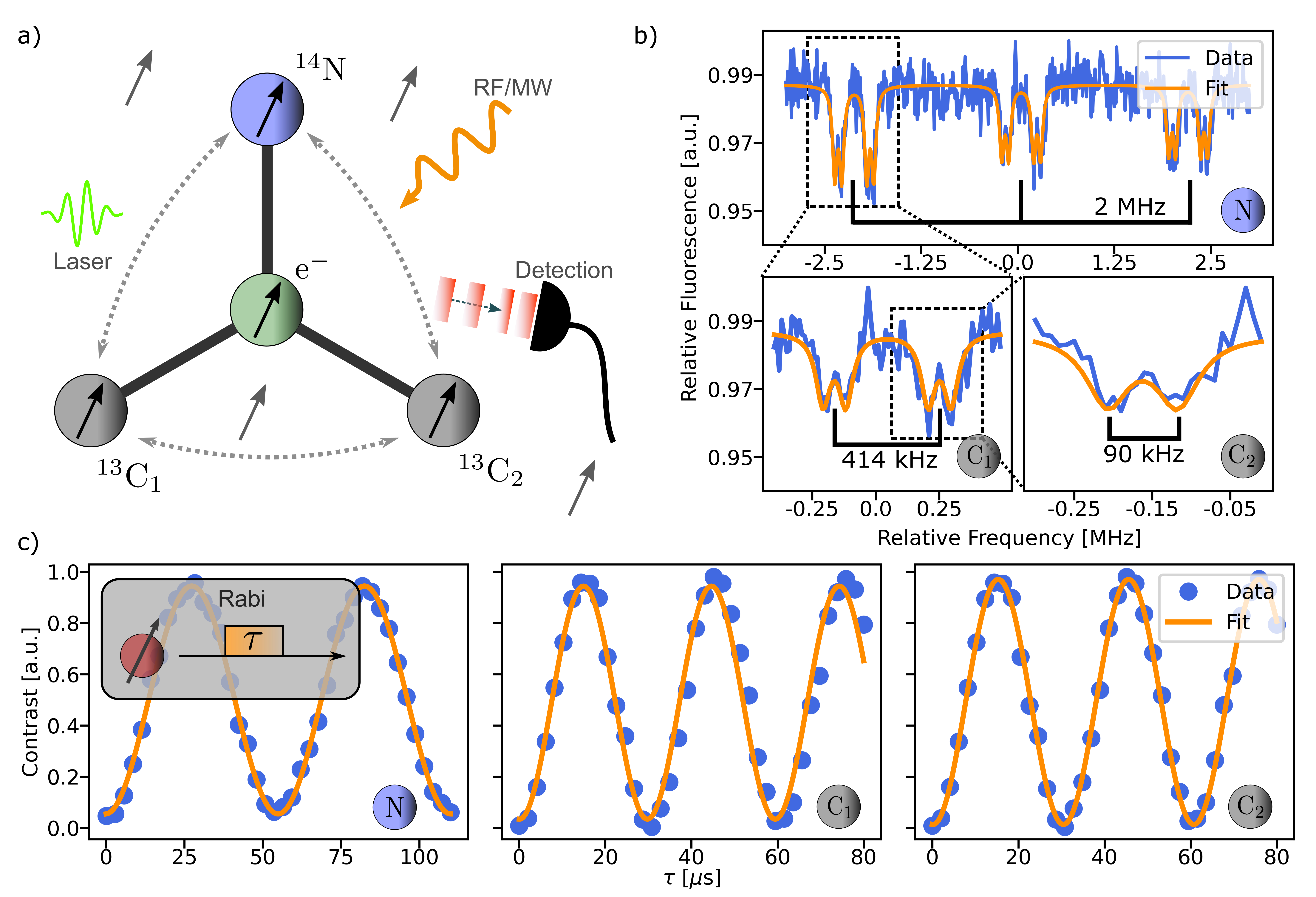}
    \caption{\textbf{a)} The electron spin of the NV center (green) is strongly coupled to three nuclear spins, which include one $^{14}$N (blue) and two $^{13}$C nuclei (grey), along with a weakly coupled spin bath composed of $^{13}$C. 
    The optical transitions of the NV center are driven by green (\SI{532}{\nano\meter}) and red (\SI{637}{\nano\meter}) lasers, whereas the charge state is detected using an orange (\SI{594}{\nano\meter}) laser. 
    The electron spin rotations and electron spin-controlled nuclear rotation can be achieved using microwave (MW) and radio frequency (RF) pulses. 
    The fluorescence, collected by a solid immersion lens (SIL), can reach up to 800 kcounts/s and is detected by an Avalanche Photodiode (APD). 
    \textbf{b)} The optically detected magnetic resonance (ODMR) spectrum shows the splitting of the nuclear spins. 
    The nitrogen splits up the $\ket{m_I=-1,0,1}$ states by \SI{2.19}{\mega\hertz}. 
    The carbon $\ket{\pm 1/2}$ states are split by \SI{414}{\kilo\hertz} for the first carbon and \SI{90}{\kilo\hertz} for the second. 
    \textbf{c)} Radio frequencies drive the nuclear spins and create Rabi oscillations between the spin states of each nuclear spin.
    Both carbon spins have a similar Rabi-frequency, which is faster compared to the nitrogen spin Rabi-frequency.} 
    \label{fig:1}
\end{figure*}
The nitrogen vacancy center (NV center) is a point defect in diamond with a substituting nitrogen atom and a vacancy. 
An electron spin of the formed defect is interacting via the hyperfine coupling with individual nuclei of $^{13}$C atoms in vicinity of the defect. 
Figure \ref{fig:1} a) displays a schematic of the formed spin register with three strongly coupled nuclear spins which are used as qubits and are driven by radio frequency (RF) and microwave (MW) pulses. 
The transition frequencies depend on the state of the electron spin of the NV center. 
In this work, the electron spin is used as an ancillary qubit and acts as a connector between the three nuclear spins. 
In addition to these strongly coupled nuclear spins, a spin bath of $^{13}$C nuclear spins is weakly coupled to the NV center. 
The electron spin is excited by an off-resonant green laser (\SI{532}{\nano\meter}), as well as a red laser (\SI{637}{\nano\meter}), which is close to the zero phonon line transition frequency between the ground and excited state and is used to enhance the initialization fidelity of the electron spin at room temperature.
The neutral charge state of the NV center is optically detected by an orange (\SI{594}{\nano\meter}) laser which is used to post-select experimental runs with the relevant NV$^-$ state. 
A microwave field (MW) is used to drive the electron spin $\ket{m_s=0 \longleftrightarrow -1,1}$ transitions, where the degeneracy is lifted by the static external magnetic field. 
Radio frequencies (RF) are employed to drive the nuclear spin transitions. With this configuration, it is possible to create a set of basis gates for the electron spin as well as for the three nuclear spins. 
The defect is operated at room temperature and has a detectable fluorescence photon count rate of up to 800 kHz, using a solid immersion lens for increased optical efficiency.
The setup is described in more detail in the methods section.
The hyperfine splitting of the nuclear spin states is visible in the optically  detected magnetic resonance (ODMR) measurement, where the frequency of microwave pulsed excitation is swept while the NV center is probed with short laser probe pulses. 
At resonance the electron spin state will flip from the bright $\ket{m_s=0}$ state to the dark $\ket{m_s=-1,1}$ state, resulting in a reduced photon count rate.%
\begin{figure*}[ht]
\includegraphics[width=\textwidth]{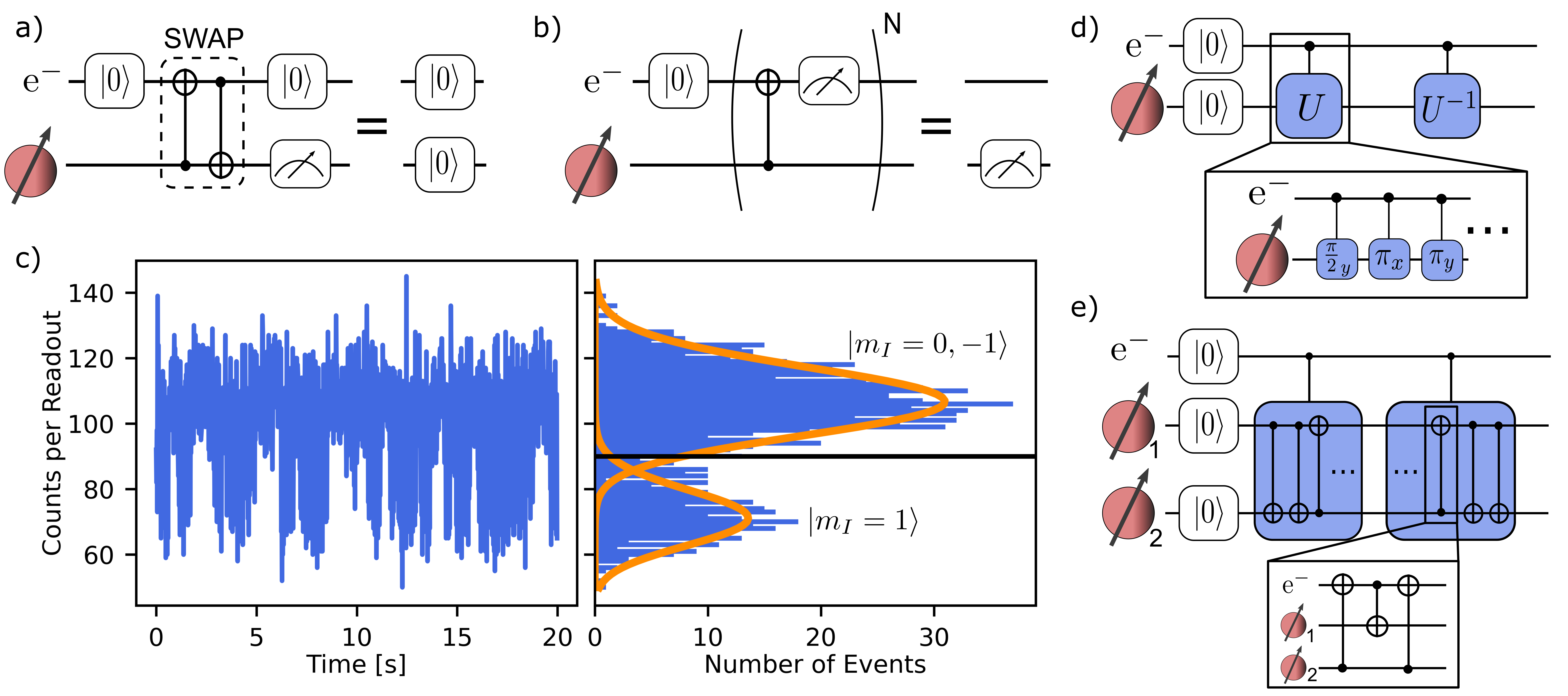}
    \caption{\textbf{a)} Nuclear spins are initialized using a SWAP gate between the pre-initialized electron spin and a nuclear spin. 
    The initialization fidelity is further enhanced by conducting a readout of the nuclear spin state after the complete initialization so that incorrectly initialized states can be discarded via post-selection. 
    \textbf{b)} The readout of the nuclear spin state is performed using a single-shot readout (SSR) which is a non-destructive projective measurement that maps the nuclear spin repeatedly onto the electron spin, which is then read out optically, re-initializing the electron spin.
    \textbf{c)} The counts provided by the SSR offer insights into the state of the nuclear spin.
    A histogram shows two Poisson distributions, that correspond to a state.
    An increased readout statistic provides a more accurate readout.
    \textbf{d)} An exemplary one-qubit randomized benchmarking (RB) circuit designed for implementation on a nuclear spin.
    After the initialization, a randomized unitary operation $U$ is applied to the nuclear spin.
    This unitary operation is controlled by the $\ket{0}$ state of the electron spin and consists of $N/2$ individual basic gates.
    Following this, the inverse unitary operation $U^{-1}$ is applied to revert the nuclear spin to the initial state.
    After a total of $N$ pulses, the probability of returning to the initial state is measured.
    \textbf{e)} Similar to single-qubit RB, the two-qubit RB is performed by applying a randomized unitary operation $U$, which is constructed from nuclear CNOT gates, followed by its inverse $U^{-1}$.}
    \label{fig:2}
\end{figure*}
Figure \ref{fig:1} b) displays the high resolution ODMR spectrum with a total of 12 resonant lines originating from $^{14}$N (2.19 MHz) and two carbon atoms $^{13}$C$_1$ (414 kHz) $^{13}$C$_2$ (90 kHz).
The $^{14}$N has a nuclear spin $I=1$, which results in three resonances, one for each $m_I=-1, 0, 1$ state. 
These transitions are separated by the hyperfine coupling to the electron spin of \SI{2.19}{\mega\hertz}. 
Each transition is further split by \SI{414}{\kilo\hertz}, corresponding to the hyperfine coupling of the $m_I=\pm 1/2$ states of the first $^{13}$C. 
Finally, a second $^{13}$C nuclear spin splits the spectrum even further with a coupling strength of \SI{90}{\kilo\hertz} to the electron spin.
In the following, the first carbon spin with a hyperfine coupling of \SI{414}{\kilo\hertz} will be refereed to as $^{13} \rm C_1$, while the second carbon spin with the \SI{90}{\kilo\hertz} coupling will be called c$^{13} \rm C_2$.
Both carbon atoms are carbon-13 isotopes and therefore have a nuclear spin $I=1/2$, resulting into two nuclear spin states each.
All three nuclear spins can be driven independently of each other. 
The Rabi oscillations for each nuclear spin are shown in figure \ref{fig:1} c). 
The nitrogen nuclear spin has the longest Rabi period $T=\SI{55}{\micro\second}$ and a contrast of \SI{0.89}{} due to the weaker $^{14}$N nitrogen gyromagnetic ratio. 
The contrast is twice the amplitude of the Rabi oscillation and describes how well a qubit can be initialized, read out and controlled.
The Rabi periods of the two carbon nuclear spins are similar, with $T=\SI{29}{\micro\second}$ and $T=\SI{30}{\micro\second}$ for $^{13} \rm C_1$ and $^{13} \rm C_2$, respectively. 
The contrast for $^{13} \rm C_2$ (0.96) is slightly higher than for $^{13} \rm C_1$ (0.91). 
This gives a total of three strongly coupled nuclear spins which can be addressed individually and are utilized as qubits in this work.
\begin{figure*}[ht]
    \includegraphics[width=\textwidth]{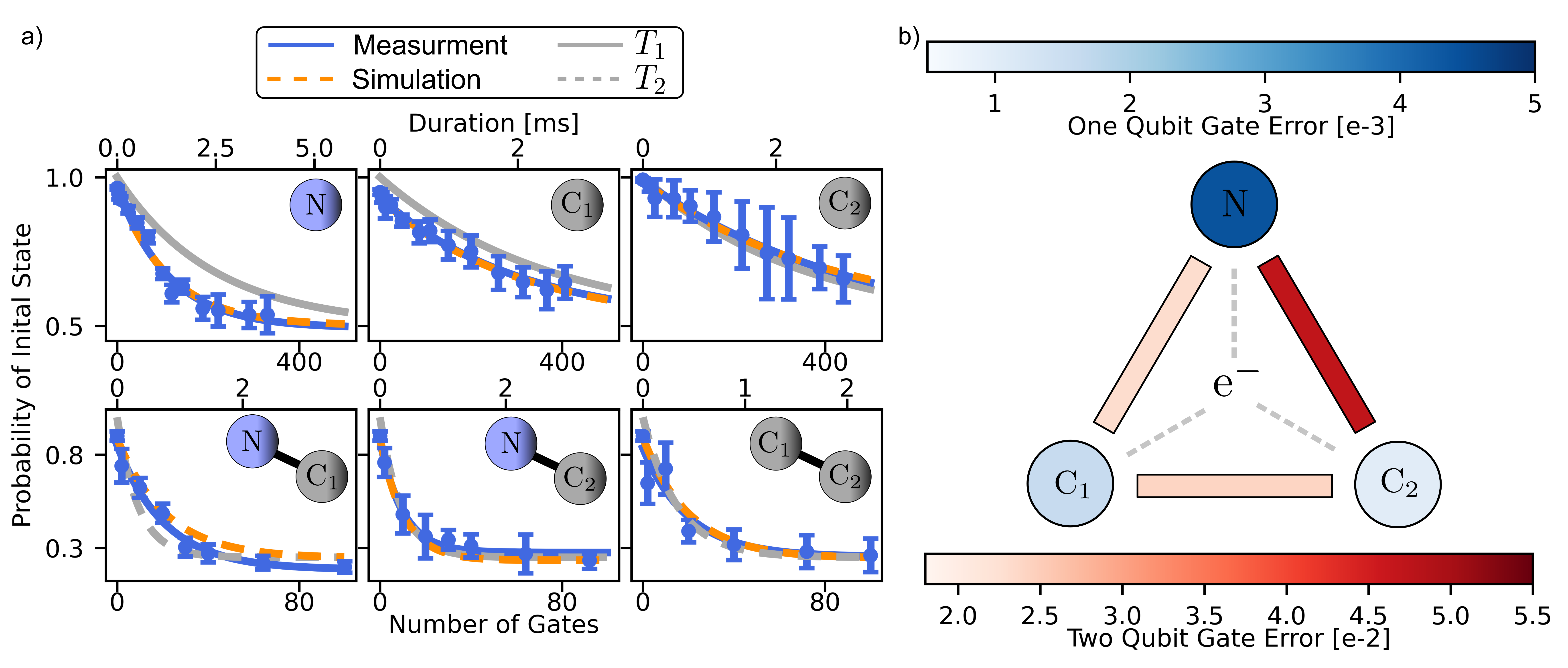}
    \caption{\textbf{a)} The benchmarking results reveal an exponential decline in the probability of returning to the initial state (blue). 
    The single-qubit benchmarking was executed on the nitrogen spin N, carbon-1 spin C$_1$, and carbon-2 spin C$_2$. 
    The two-qubit randomized benchmarking was performed across all three possible qubit connections. The errors of the single-qubit gates are limited by the $T_1$ of the electron spin, while the two-qubit gates are limited by the $T_2$. The simulation, that was extracted from benchmarking fit all benchmarking results.
    \textbf{b)} The measured errors of the basic gates can be visualized through an error map of the system. 
    This error layout was subsequently used to construct a noise model of the system. }
    \label{fig:3}
\end{figure*}
To initialize a nuclear spin, the electron spin has to be initialized first, using the green off-resonant laser pulse to pump the NV center into its negatively charged state and on-resonant lasers to cycle the electron spin in the desired $\ket{m_s = 0}$ state. 
Two nuclear spin state selective microwave pulses yield a controlled NOT gate on the electron spin and can be used to transfer the initialized electron spin state to the nuclear spin, while the state of the electron spin is re-initialized after this sequence in order to prepare it for subsequent nuclear spin readout.
This process is illustrated in figure \ref{fig:2} a).
To enhance the initialization fidelity of the state, we use the method of post-selection, where all incorrectly initialized states are discarded. 
 The readout of a nuclear spin is done via a single-shot readout (SSR) \cite{neumann}, as illustrated in figure \ref{fig:2} b). 
For this, the nuclear spin state is projected onto the initialized electron spin, which is then read out.
The once projected state of the nuclear spin is not disturbed by the operation of the projection, nor by the measurement of the electron spin. 
This allows multiple readouts of the electron spin, which makes this process a quantum non-demolition (QND) measurement. 
This readout procedure is conducted $N$ times to enhance the statistical significance of the collected data and to accurately extract the projection of the nuclear spin state along the z-axis. 
When the nuclear spin is in state $m_I=1$, addressed by the selective MW pulse, upon readout the defect yields fewer photons, which can be seen in figure \ref{fig:2} c).
The fluorescence time trace over multiple readout processes depicts the spin-flips of the nuclear spin memory in real-time. 
When the time trace is displayed as a histogram, the Poisson distributions of the two states are clearly distinguishable, thus enabling single-shot state estimation via the maximum likelihood method \cite{ZahedianOnReadout}. 
For optimal maximum likelihood estimation, a photon number threshold is set to classify the measured states.
If the number of counts is above this threshold, the state is attributed to $m_I=\{0,-1\}$, if the state is below the threshold, the state can be identified as the $m_I=1$ state.
By setting the threshold higher during the postelection process, the readout fidelity can be boosted, since the error of overlapping Poisson distributions can be minimized with a natural trade-off of lower initialization success rate and larger time overhead \cite{ZahedianOnReadout}.
\section{Benchmarking}
To characterize the fidelity of our quantum processor, we used the method of randomized benchmarking (RB), which involves the application of Clifford unitary operations with a circuit mirroring technique \cite{Proceedings}. 
\subsection{Randomized Benchmarking}
For characterizing the error per gate (EPG) of quantum operations, randomized benchmarking has several advantages compared to quantum process tomography. 
It requires a number of measurements that scales polynomially with the number of qubits to determine the EPGs.
This is helpful to benchmark large quantum systems since quantum process tomography scales exponential with the number of qubits. 
Since the EPG is extracted through the exponential decay of the state fidelity as a function of the number of gates in a sequence, the infidelity of the state initialization and readout does not affect the extracted EPG.
This extracted EPG helps to evaluate the performances of lengthy quantum operation sequences, which is relevant for quantum information processing algorithm applications. 
In this work, we demonstrate randomized benchmarking on qubits in a single NV color center system. 
The experiment characterizes how the probability of measuring the initial state evolves for a single qubit or two qubits under the applied Clifford gate sequences. 
In a randomized benchmark circuit, a number $N$ of random Clifford gates are applied, forming an unitary operation $U$. Afterwards, the inverse of the sequence $U^{-1}$ is applied, using the reversibility of quantum computing to reproduce the initial state. 
These Clifford gates need to be transpiled into the set of basis gates which are native on the quantum processor. On the NV centers, native basis gates are single-qubit rotational gates around the $x$ and $y$ axis on the Bloch sphere as well as controlled two-qubit rotations of either the electron conditioned on the state of the nuclear spin ($C_nNOT_e$) or the nuclear spins conditioned on the state of the electron spin ($C_eNOT_n$).
Due to errors - whose main sources are spin relaxation and dephasing, cross talk, fluctuations in the microwave control field - the probability of measuring the same initial state after the application of $N$ rotational gates will decrease with $N$.
The error probability of each gate is multiplied with each other to obtain the total error of the quantum circuit.
Due to this multiplication, the error of the circuit scales exponential with $N$, until the states cannot be distinguished anymore.
This decrease of the initial state probability  is modeled by
\begin{equation} 
F = A_0\cdot \alpha ^N + B_0.
\label{eq:fit}
\end{equation}
By fitting this curve to the experimental data, the parameter $\alpha$ can be determined \cite{formula}, which is used to extract the error per gate via
\begin{equation}
    \text{EPG}(n,\alpha) = \frac{2^n-1}{2^n}\cdot (1- \alpha).
\end{equation}
This error also depends on the number of qubits $n$, on which the gates were performed. 
State initialization and readout errors are represented by the fitting parameters $A_0$ and $B_0$, which do not influence the EPG.
Readout and initialization are only applied once for each circuit, therefore its influence on the overall error is constant and does not scale with $N$.
An exemplary circuit for a one- and a two-qubit randomized benchmark is displayed in figure \ref{fig:2} d) and e), respectively. 
In the one qubit benchmark randomized native rotations are applied on a nuclear spin. These operations are mirrored to obtain the initial state, that is then read out.
The benchmark for a two-qubit controlled-NOT operation was performed in an equivalent manner to the benchmarking of the single-qubit gates, where instead of single-qubit gates two-qubit CNOT gates were applied. 
Since the coupling between two nuclear spins is too weak to perform efficient gates, three CNOT gates between the nuclear spins and the electron spin are applied to achieve an operation that is logically equivalent to a CNOT between two arbitrary nuclear spins.
The first CNOT pulse projects the state of the control nuclear spin qubit onto the electron spin through rotation. 
This is followed by a CNOT operation between the electron spin and the target nuclear spin of the logical sequence, where the electron spin is the control qubit and the nuclear spin the target. 
With the third pulse of the sequence, the projection of the first nuclear spin on the electron spin is reversed. 
\subsection{Results}
The results of the randomized benchmarks are illustrated in figure \ref{fig:3} a). 
Since the $^{14}N$ nuclear spin has the longest Rabi period, the benchmarking circuits take longer to execute, which induces a larger error per gate due to the spin relaxation of the electron spin. 
The exponential decay gives an EPG of $4.4(2)\cdot 10^{-3}$ for $^{14}\rm N$ spin. 
The two carbon nuclear spins perform better than the $^{14}\rm N$ spin with an EPG of $1.6(3)\cdot 10^{-3}$ and $1.0(5)\cdot 10^{-3}$ for $^{13} \rm C_1$ and $^{13} \rm C_2$, respectively.
This is mostly attributed to the faster Rabi oscillation of the carbon spins, allowing more gates within the longitudinal relaxation time $T_1=\SI{2.22}{\milli\second}$ of the electron spin. 
For the NV center, the coherence time of the central electron spin is much smaller than the coherence time of the surrounding nuclear spins ($>\SI{10}{\milli\second}$).
All rotations, performed on the nuclear spins, are depending on the state of the electron spin.
Hence a $T_1$ induced error on the electron spin decreases the fidelity of the nuclear spin gate.
Since the electron spin is preserved in its eigenstate the $T_2$ induced errors will not contribute. 
The $T_1$ time of the electron spin is the limitation of the single qubit gates, indicating that further improvement of the gate fidelity at room temperature is not possible and requires cooling down to cryogenic temperatures \cite{AcostaT1}.
In comparison to $^{13} \rm C_1$ and the $^{14} \rm N$ spin, the measurement values of the randomized benchmark of $^{13} \rm C_2$ show a larger $\sigma$ deviation of the data.
Due to the comparatively weak coupling between $^{13} \rm C_2$ and the electron spin of $\SI{90}{kHz}$, addressing this nuclear spin requires a high stability of the transition frequency.
\begin{figure*}[ht]
    \centering
    \includegraphics[width=\textwidth]{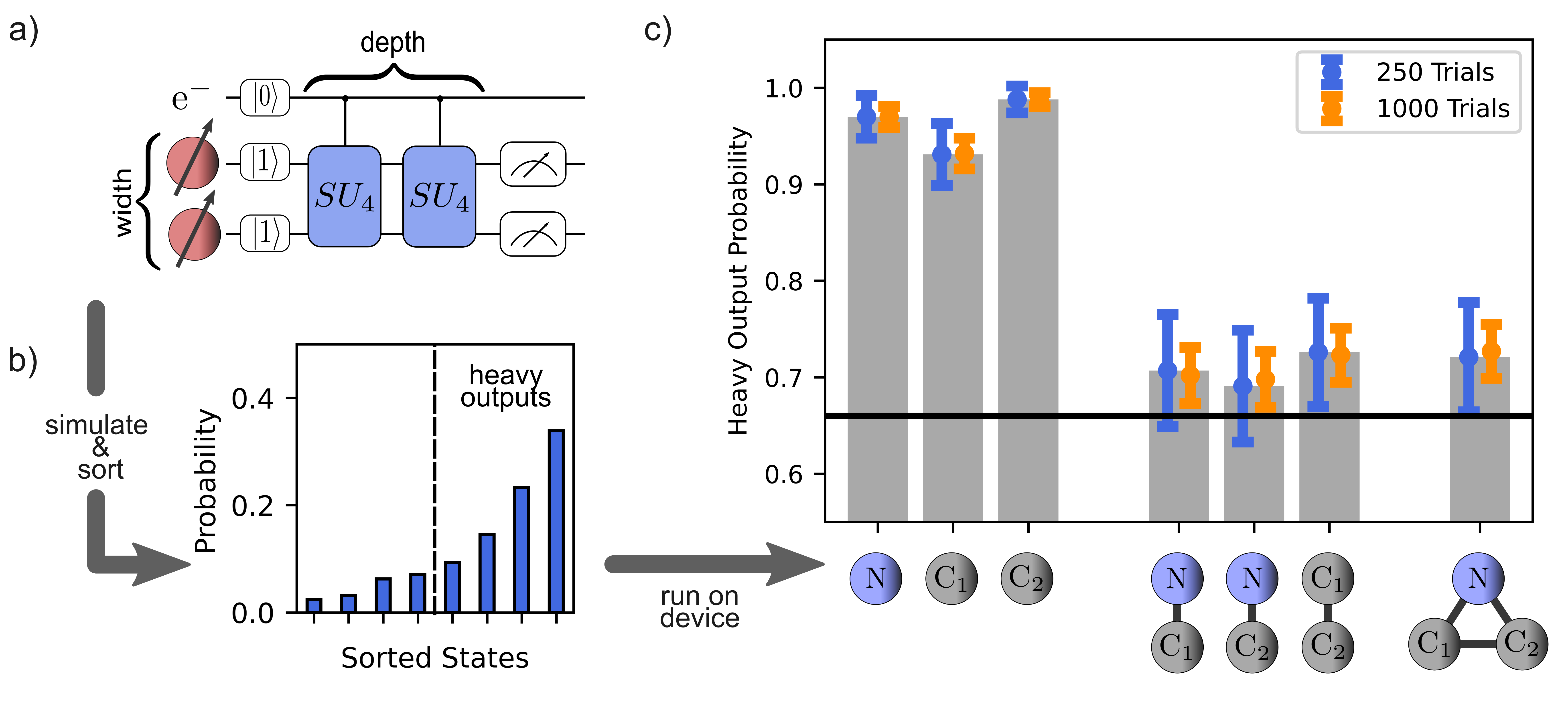}
    \caption{\textbf{a)} Quantum volume circuits are constructed using $SU(4)$ matrices. The number of qubits involved is referred to as the width, while the quantity of $SU(4)$ matrices is known as the depth of the circuit. 
    The quantum volume is determined by the maximum width and depth of a successful circuit.
    \textbf{b)} To evaluate whether a circuit is successful, one must simulate it and sort the readout states based on their probabilities. 
    States above the median are referred to as "heavy outputs". 
    If the state probability of all heavy outputs combined, within the $2\sigma$ deviation, from at least 100 trials exceeds 0.66, the circuit is considered successful. 
    \textbf{c)} The heavy output probability for a quantum volume measurements is simulated across all individual qubits and their connections. 
    After a minimum of 100 trials, the measurements exceeded the threshold, achieving a maximum width and depth of $m = 3$, which corresponds to a quantum volume of $2^m = 8$.}
    \label{fig:4}
\end{figure*}
The errors seen in the experiment are induced by a shift of the transition frequency originating from heating effects related to MW and RF pulses during the experiment.
The noise model, created from the EPG, the readout error, and the initialization error, reproduces the measurement results of the nuclear spins very well.
The noise model was created analog to \cite{Finsterhoelzl_2023, qiskit2024}.
This shows that it is possible to extract all relevant single-qubit errors from single-qubit randomized benchmarks and use them to create a single-qubit model of the quantum register. 
Two qubit gates are a crucial part of a quantum computing algorithms, because they are used for the essential step of entangling two qubits.
High qubit connectivity, in systems such as nitrogen-vacancy (NV)–based networks or trapped ions, facilitates easy control over two-qubit couplings for two-qubit gates. 
Effectively connecting two qubits through ancilla qubit-mediated interactions enhances the fidelity of two-qubit gate operations, which are fundamental for building multi-qubit gates. 
This advancement is essential for applications in quantum computing and quantum networks\cite{PhysRevA.82.020301}, \cite{ruf2021quantum}, \cite{pnas.2314103121}.
The two-qubit gate benchmark was conducted on all three possible qubit combinations, as shown in figure \ref{fig:3} a). 
The benchmark between the $^{14}\rm N$ and $^{13} \rm C_1$, as well as $^{13} \rm C_1$ and $^{13} \rm C_2$ show a similar performance with an EPG of $2.3(1)\cdot 10^{-2}$ and $2.4(4)\cdot 10^{-2}$, respectively. 
The benchmark between the $^{14} \rm N$ spin and $^{13} \rm C_2$ spin performs the worst with an EPG of $4.7(4)\cdot 10^{-2}$. 
One possible explanation is a combination of the slower Rabi rate of the $^{14} \rm N$ spin and the heating-induced frequency shifts influencing $^{13} \rm C_2$. 
The benchmarks show, that the two qubit gates are limited by the $T_2$ of the electron spin. 
The observation indicates that two-qubit randomized benchmarking reveals an increased susceptibility to decoherence and errors in the experiment sequence compared to the single-qubit randomized benchmarking \cite{Fidelity}. 
Once again the numerical results obtained with the noisy simulator created from the fit parameters are in accordance with the measurement.
Small differences between the measurement and the noise model originate from the deviations of the simulation and the measurement, due to limited data.
All individual EPG are shown in the error map of the system, illustrated in figure \ref{fig:3} b). 
The corresponding error map and connectivity are used in the noise simulation model. 
\subsection{Extracting Quantum Volume}
Quantum volume is a metric to measure the error characteristics of a general quantum hardware platform. 
It is obtained by implementing a randomized circuit and evaluating the results.
It provides a single numerical value that serves as a measure to compare the capabilities of different platforms for quantum computation. 
In a quantum volume measurement, all errors of the quantum device are tested in the context of an overall performance, since a large number of gates are performed on and in between all qubits. 
Notably, the quantum volume of a system is decreased by uncontrolled interactions within the system during the experiments.
Random circuits are broadly applied to benchmark, quantum computing platforms \cite{Characterizing}.
Quantum volume quantifies the largest square-shaped random circuit of equal width (the number of qubits involved) and depth (the number of operations) $m$ that it can successfully run \cite{Quantum_Volume}, \cite{Moll_2018}. 
A run is defined as successful when the average heavy output probability as well as its 2$\sigma$ deviation is fully above a 2/3 threshold.
The heavy output probability is defined as the combined probability of detecting any state, that occurs with higher probability than the median value \cite{quantumsupremacy}, as indicated in figure \ref{fig:4} b). 
The heavy output probability constitutes a reliable quantitative marker of the quantum circuit fidelity because any noise impedes the ideal uniform distribution of output probabilities.
The number of trials can be increased indefinitely.
The quantum volume $V_Q$ is then defined as 
\begin{equation}
    V_Q = 2^m \thickspace .
\end{equation}
To test the simulation of a quantum volume, a quantum volume circuit is designed for a specified number of qubits which is composed of multiple quantum layers. 
Within each quantum layer, unitary operations are applied to randomly chosen qubit pairs. 
The unitary operations, applied to each qubit pair, are described by $SU(4)$ matrices, ensuring an equal likelihood of obtaining any quantum state \cite{PhysRevA.100.032328}. 
Such a circuit for width and depth of $m = 2$ is shown in figure \ref{fig:4} a).
If the qubits in a quantum device are fully depolarized in a quantum circuit sequence, the expected heavy output probability reduces to 0.5, while a quantum circuit with fully polarized qubits achieves a heavy output probability of 1.0.
The key objective of heavy output generation is to create a collection of output strings (the resulting quantum states) where a majority exceeding two-thirds are classified as heavy \cite{quantumsupremacy}.
Our simulations for quantum volume circuits are performed with the noisy model, that was created based on the results of the randomized benchmarking \cite{qiskit2024}.
The results of the quantum volume simulation are given in figure \ref{fig:4} c). 
To establish a robust statistical confidence in exceeding the threshold of 2/3, the simulation protocol was repeated with 1000 circuit instances, while seed 42 was used \cite{qiskit2024}. 
The quantum volume circuit for $m=3$ was successful on all possible qubit connections, achieving a quantum volume of 8, which is the maximum for a system consisting of three nuclear qubits. 
This demonstrates that the three-qubit room-temperature NV center presented in this work is a viable platform for quantum computing.
All three strongly coupled nuclear spins can be used for quantum computation, within the limit of a quantum volume measurement which marks the record quantum volume metric for ambient condition operation to date.
A quantum volume of 8 in a room-temperature NV center setup, limited by the system's $T_1$ and $T_2$ times, suggests that low-temperature experiments could further enhance the benchmarking results. 
This indicates that low-temperature NV centers hold significant potential as quantum network nodes with integrated quantum computing capabilities \cite{ruf2021quantum}.
The noise model developed in this work can also be used to simulate and predict more complex experiments that would be time-consuming on the NV center, as the heating from lengthy sequences can shift the transition frequencies.
Measurements exhibiting a shift in the ODMR frequency of \SI{15}{\kilo\hertz} will be discarded, which contributes to the extended measurement duration. 
\begin{figure*}[ht]
    \centering
    \includegraphics[width=0.9\textwidth]{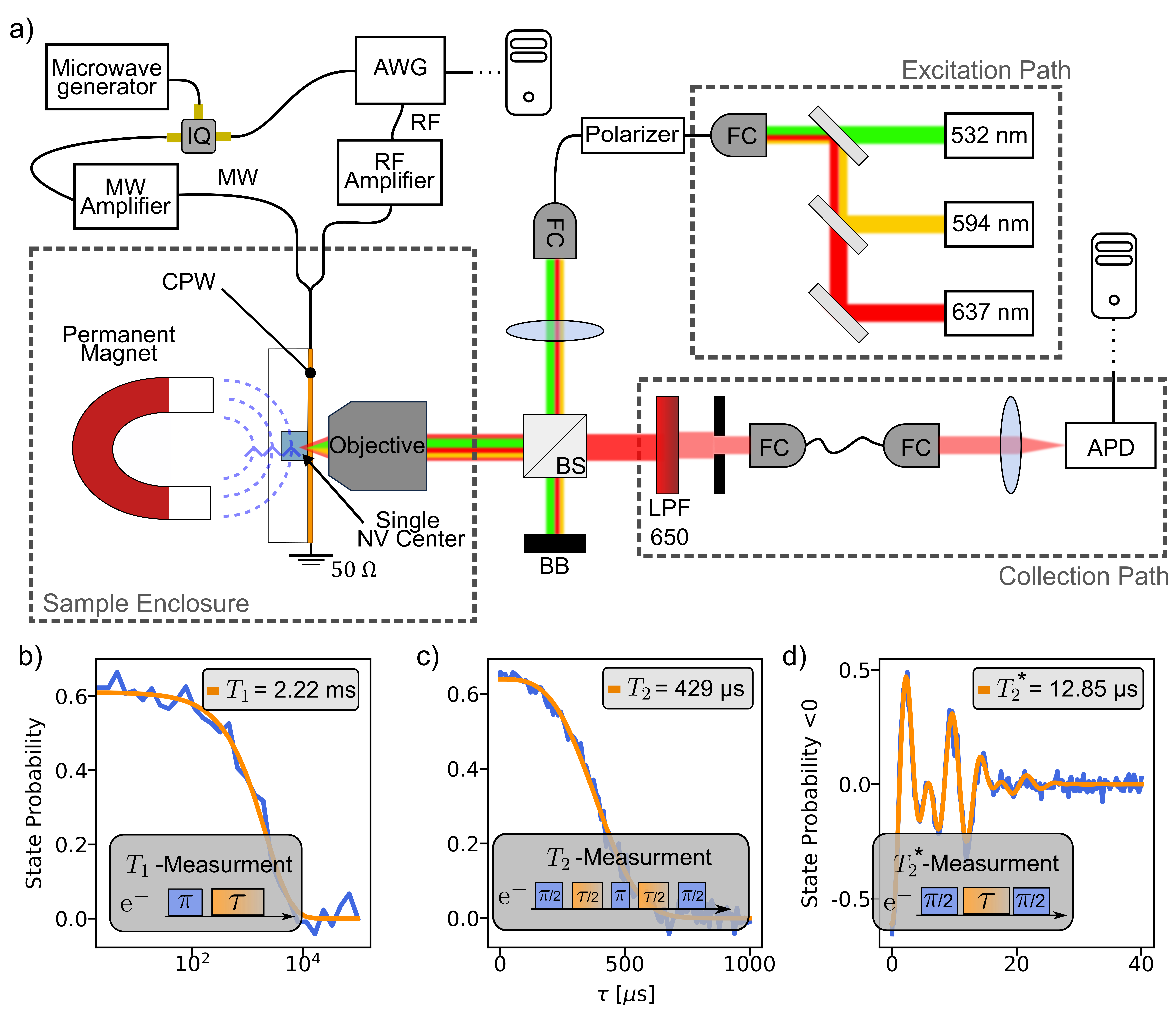}
    \caption{\textbf{a)} Schematic of the experimental setup. 
    The pulse sequences are generated by the PC and executed by the arbitrary waveform generator (AWG). 
    The laser pulses are combined and delivered to the NV center, which is close to a permanent magnet, by an oil-immersion objective. 
    The fluorescence is collected by the same objective, filtered by a pinhole and a longpass filter, and detected by an APD. 
    These counts are processed by the PC. 
    The microwave (MW) and radio frequency (RF) pulses, generated by the AWG and microwave generator, are delivered to the sample via a gold strip on the diamond surface. 
    \textbf{b)} $T_1$-, \textbf{c)} $T_2$-, and \textbf{d)} $T_2^*$-measurement, performed on the electron spin of the NV center. 
    The $T_2^*$ measurement displays two oscillations originating from the two strongly coupled carbon spins.}
    \label{fig:setup}
\end{figure*}
\section{Methods}
Figure \ref{fig:setup} a) displays a sketch of the setup. 
The \SI{532}{\nano\meter} laser is used to repump and initialize the electron spin of the NV center.
The yellow \SI{594}{\nano\meter} laser is used to detect and filter the neutral charge state of the NV center (NV$^0$).
Analogous to the single-shot readout described in figure \ref{fig:2}, the two charge states can be distinguished.
As in the single-shot readout, the measurements that read the NV$^0$ state are discarded in the post selection.
The red \SI{637}{\nano\meter} laser is resonant to the direct optical transitions of the NV center and is used to enhance the initialization fidelity.
The yellow and green lasers are home-built diode lasers, while the red laser is a tunable diode laser from Toptica.
All lasers are combined by matching dichroic mirrors and are coupled into an optical fiber.
A polarizer adjusts the polarization of the laser light in the optical fiber.
Afterward, the light is coupled out and is directed to the immersion oil objective using a beamsplitter.
The mounted diamond sample, containing the single NV center, is located in the focus of the objective.
The sample is mounted onto positioners, such that the position of the diamond can be aligned relative to the objective.
The permanent magnet is also mounted on a positioner, this makes it possible to align the magnet to the NV center and therefore alter the transitions, affected by Zeeman splitting.
During the measurements, the NV center experienced a magnetic field of around \SI{0.6}{\tesla}.
The fluorescence of the NV center is collected by the same immersion oil objective and is transmitted through the beamsplitter to a pinhole.
This pinhole is used to filter out noise and ensure that only light of the NV center in focus is collected.
A \SI{650}{\nano\meter} longpass filter is used to stop laser light that is reflected off the diamond surface.
To reduce dark counts, the resulting phonon sideband is coupled into an optical fiber that leads to an enclosure shielded from environmental light.
The fluorescence, collected by the APD, is evaluated by a Time Tagger from Swabian Instruments, which provides the number of counts to the PC.
This PC is also responsible for creating the pulse sequences that involve lasers, radio frequencies (RF), and microwaves (MW).
This pulse sequence is then sent to the arbitrary waveform generator (AWG), which plays the desired sequence.
In the experimental setup, we use the Keysight Agilent M8190a as an AWG.
The AWG is able to create RF up to \SI{100}{\mega\hertz}.
This is sufficient for the RF pulses of the nuclear spins.
For the creation of the MW pulses, we use an additional microwave generator (Anritsu MG3702xA) that constantly creates a set MW frequency.
To sweep the frequency, the constant MW is combined at an IQ-Mixer with the RF output of the AWG. 
The frequencies of the AWG are added to the MW frequency, making it possible to sweep and pulse the MW.
The resulting MW and RF pulses are amplified using separate amplifiers and then combined. 
For RF signals, we use the AR 500A250D, while the Gigatronics GT-1000A is employed for amplifying MW sequences.
These pulses are applied to the NV center by a gold strip that is deposited on the diamond surface. 
This golden strip strop forms a coplanar waveguide (CPW) that acts as an antenna between the signal line and the surrounding ground plane.
The pulses are led away from the sample and are damped into a heat-sinked \SI{50}{\ohm} terminator.
This termination is done away from the sample to reduce heating effects at the diamond sample.

The $T_1$, $T_2$, and $T_2^*$ measurements of the electron spin, along with their sequences, are shown in figure \ref{fig:setup} b)-d).
The $T_1$ relaxation time is measured at \SI{2.22}{\milli\second} while the Hahn Echo $T_2$ is \SI{429}{\micro\second}.
The $T_2^*$ measurement, displayed in figure \ref{fig:setup} d), shows two oscillations.
These oscillations originate from the two strongly coupled carbon spins.
The nitrogen spin does not contribute to the oscillations, as it was initialized beforehand.
The data can be fitted using following function
\begin{equation}
    Y(\tau) = A\cdot\exp\biggl(-\frac{\tau}{T_2^*}\biggr)\cos(\tau f_1)\cos(\tau f_2).
\end{equation}
Here $f_1$ and $f_2$ correspond to $A_{zz}/2$ for two strongly coupled carbon spins.
The $A_{zz}$ coupling of the strongly coupled nuclear spins to the electron spin, along with all other characteristics of the strongly coupled nuclear spins and the electron spin, are displayed in table \ref{tab:1}. 
These parameters were determined using the ODMR sequence, shown in figure \ref{fig:1}. 
The Rabi key features are extracted from the Rabi oscillations shown in figure \ref{fig:1}.  
The $T_1$, $T_2$, and $T_2^*$ parameters of the spins are also displayed. 
The $T_2$ of the nuclear spins is limited by the electron $T_1$ time, as a bit flip on the electron spin induces decoherence at the rate of the hyperfine coupling \cite{zaiser}.
\begin{table*}[ht]
    \centering
     \caption{Detailed characteristics of the individual spins in the system. 
     The $A_{zz}$ parameter describes the strength of the coupling term in the Hamiltonian. 
     $T_1$, $T_2$, and $T_2^*$ represent the spin relaxation and dephasing characteristics. 
     The Rabi contrast and period characterize the MW and RF drives of the nuclear spins. 
     The one qubit error per gate (1-Qubit EPG) and the two qubit error per gate (2-Qubit EPG) were measured using randomized benchmarking.}
    \begin{tabular}{c||c|c|c|c}
        Spin  &electron spin & $^{14}$N & $^{13}$C$_1$ & $^{13}$C$_2$ \\ \hline\hline
        $A_{zz}$ & - & \SI{2}{\mega\hertz}&\SI{414}{\kilo\hertz} &\SI{90}{\kilo\hertz} \\ \hline
        Rabi period & \SI{1.10}{\micro\second}&\SI{55.13}{\micro\second} & \SI{29.69}{\micro\second} & \SI{30.28}{\micro\second} \\ \hline
        Rabi contrast & &0.890 & 0.909 & 0.956 \\ \hline
        $T_1$ & \SI{2.22}{\milli\second}& $> \SI{100}{\second}$ \cite{zaiser}& $> \SI{100}{\second}$ \cite{zaiser}&$ > \SI{100}{\second}$\cite{zaiser}\\ \hline
        $T_2$ & \SI{429}{\micro\second}&$\approx$ 1.5$T_{1,e}$ \cite{zaiser}& $\approx$ 1.5$T_{1,e}$  \cite{zaiser}& $\approx$ 1.5$T_{1,e}$  \cite{zaiser}\\ \hline
        $T_2^*$ & \SI{12.85}{\micro\second}&-&-& -\\ \hline
        1 Qubit Epg & - & $(4.4\pm0.2)\cdot 10^{-3}$ & $(1.6\pm0.3)\cdot 10^{-3}$ & $(1.0\pm0.5)\cdot 10^{-3}$ \\ 
    \end{tabular}
    \\
    \bigskip
    \begin{tabular}{c||c|c|c} 
        Qubit con. & $\text{N}-\text{C}_1$ & $\text{N}-\text{C}_2$ & $\text{C}_1-\text{C}_2$ \\ \hline \hline
        2 Qubit Epg &   $(23\pm 1)\cdot 10^{-3}$ & $(47\pm 4)\cdot 10^{-3}$ & $(24\pm 4)\cdot 10^{-3}$ \\
    \end{tabular}
    \label{tab:1}
\end{table*}
In addition to the three strongly coupled nuclear spins discussed above, the electron spin is also coupled to a nuclear spin bath of $^{13}$C nuclear spins.
These weakly coupled nuclear spins can be sensed using the electron nuclear double resonance (ENDOR) sequence \cite{neumann}, \cite{ENDOR}. 
These nuclear spins could potentially be used to increase the quantum volume of the NV center. 
Even an added qubit assumed to have perfect initialization, perfect readout, and no errors on the quantum gates will not improve the quantum volume measurement.
This shows that the quantum volume measurement is limited by the number of $SU(4)$ operations that are applied to the three strongly coupled nuclear spins.
Weakly coupled nuclear spins cannot improve the quantum volume results due to the $T_1$ and $T_2$ limitations of the room-temperature NV center. 
However, weakly coupled nuclear spins could potentially be used as a quantum memory, to store quantum states between circuits.
\section*{Acknowledgments}
This research was supported by funding from the Land Baden-Württemberg through the projects QC4BW and KQCB24, the European Union under the SPINUS project, and the German Federal Ministry of Education and Research (Bundesministerium für Bildung und Forschung) through the QR.N initiative.
\newpage
\bibliographystyle{unsrt}
\bibliography{literature}

\end{document}